\documentclass[useAMS,usenatbib,letters]{mn2e}
\usepackage{aas_macros,graphicx,times,multirow}
\title[Quiet but still bright: \xmm\ observations of \src]
  {Quiet but still bright: \xmm\ observations of the soft gamma-ray repeater \src\thanks{Based on observations obtained with \emph{XMM-Newton}, an ESA science mission with instruments and contributions directly funded by ESA Member States and
  NASA.}}
\author[A.~Tiengo et al.]
{A.~Tiengo,$^{1}$\thanks{E-mail: tiengo@iasf-milano.inaf.it}
P.~Esposito,$^{1,2}$ S.~Mereghetti,$^{1}$ G.~L.~Israel,$^{3}$ L.~Stella,$^{3}$ R.~Turolla,$^{4,5}$
\newauthor S.~Zane,$^{5}$ N.~Rea,$^{6}$ D.~G\"otz,$^{7}$ and M.~Feroci$^{8}$\smallskip\\
$^1$INAF/Istituto di Astrofisica Spaziale e Fisica Cosmica - Milano, via E.~Bassini 15, 20133 Milano, Italy\\
$^2$INFN - Istituto Nazionale di Fisica Nucleare, Sezione di Pavia,
via A.~Bassi 6, 27100 Pavia, Italy\\
$^3$INAF/Osservatorio Astronomico di Roma, via Frascati 33, 00040 Monteporzio Catone, Italy\\
$^4$Universit\`a degli Studi di Padova, Dipartimento di Fisica, via F.~Marzolo 8, 35131 Padova, Italy\\
$^5$University College London, Mullard Space Science Laboratory, Holmbury St. Mary, Dorking, Surrey RH5 6NT, UK\\
$^6$University of Amsterdam, Astronomical Institute Anton Pannekoek, Kruislaan 403, 1098~SJ Amsterdam, The Netherlands\\
$^7$CEA Saclay, DSM/Irfu/Service d'Astrophysique, Orme des Merisiers, B\^at. 709, 91191 Gif sur Yvette, France\\
$^8$INAF/Istituto di Astrofisica Spaziale e Fisica Cosmica - Roma,
via Fosso del Cavaliere 100, 00133 Roma, Italy}

\date{Accepted 2009 July 25. Received 2009 July 24; in original form 2009 June 11}
\pagerange{\pageref{firstpage}--\pageref{lastpage}} \pubyear{2008}
\def\LaTeX{L\kern-.36em\raise.3ex\hbox{a}\kern-.15em
    T\kern-.1667em\lower.7ex\hbox{E}\kern-.125emX}

\def\xmm {\emph{XMM-Newton}}
\def\cha {\emph{Chandra}}

\def\src {SGR\,0526--66}
\def\sgr {SGR\,0526--66}
\def\flux {\mbox{erg cm$^{-2}$ s$^{-1}$}}
\def\lum {\mbox{erg s$^{-1}$}}
\def\nh {$N_{\rm H}$ }

\begin{document}

\label{firstpage}

\maketitle

\begin{abstract}
\sgr\ was the first soft gamma-ray repeater (SGR) from which a giant
flare was detected in March 1979, suggesting the existence of
magnetars, i.e. neutron stars powered by the decay of their
extremely strong magnetic field. Since then, very little information
has been obtained on this object, mainly because it has been
burst-inactive since 1983 and the study of its persistent X--ray
emission has been hampered by its large distance and its location in
a X--ray bright supernova remnant in the Large Magellanic Cloud.
Here we report on a comprehensive analysis of all the available
\xmm\ observations of \src. In particular, thanks to a deep
observation taken in 2007, we measured its pulsation period
($P=8.0544\pm0.0002$ s) 6 years after its latest detection by \cha.
This allowed us to detect for the first time a significant reduction
of its spin-down rate. From a comparison with two shorter \xmm\
observations performed in 2000 and 2001, we found no significant
changes in the spectrum, which is well modelled by an absorbed
power-law with \nh\ = 4.6$^{+0.7}_{-0.5}$ $\times 10^{21}$ cm$^{-2}$
and $\Gamma$=3.27$^{+0.07}_{-0.04}$. The high luminosity
($\sim$4$\times$10$^{35}$ \lum, in the 1--10 keV energy band) still
observed $\sim$25 years after the latest detection of bursting
activity places \src\ in the group of bright and persistent magnetar
candidates.

\end{abstract}

\section{Introduction}
On 1979 March 5, an extremely bright gamma-ray burst (GRB), followed
by a $>$60 s long tail pulsating at a period of $8.1\pm0.1$ s, was
detected by many spacecrafts \citep{mazets79}. The event was
localized within the young ($\sim$5,000--10,000 years old) supernova
remnant (SNR) LHA 120--N49 (N49) in the Large Magellanic Cloud (LMC;
\citealt{cline82}). These properties indicated that the burst was
emitted by a young neutron star, leading \citet{duncan92} and
\citet{paczynski92} to propose the existence of neutron stars with
magnetic fields of $\sim$10$^{15}$ G, that were called {\it
magnetars}. The detection of many weaker bursts from the same
direction  in the following 4 years indicated that the March 5 event
was not a typical GRB,
% (most of which are now recognized to be at cosmological distances)
but an exceptional outburst from a small class of sources which had
been just discovered and called soft gamma-ray repeaters (SGRs).
Indeed the 1979 March 5 event from \src\ was the first ``giant
flare'' observed from a SGR. Only two other such events have been
observed in the following years, each one from a different SGR
\citep{hurley99,hurley05short}.

%burst (called {\it giant flare}) from a soft gamma-ray repeater
%(SGR), that was named \src.

Up to now, only six SGRs have been discovered (plus a few
candidates). They are characterised by the emission of short bursts
of gamma-rays during sporadic periods of activity. In addition, they
are also observed as pulsating  X-ray sources with periods in the
2--9 s range and persistent luminosities up to $\sim$$10^{36}$ \lum.
The magnetar model was developed to explain both their bursting and
persistent emission \citep{thompson95} and later extended
\citep{thompson96} to the interpretation of the anomalous X-ray
pulsars (AXPs). These are a group of 9 X-ray sources (plus some
candidates) with similar properties to those of the SGRs (see
\citealt{mereghetti08} for a review).

The persistent X-ray emission from \src\ was first detected with
{\it ROSAT} \citep{rothschild94} and then observed by \cha\ in 2000
and 2001 \citep{kulkarni03,park03}, with a constant X-ray luminosity
of $\sim$$10^{36}$ \lum\ (unabsorbed, in the 0.5--10 keV energy
range). Being \src\ a rather faint X--ray source, the pulsation of
its persistent emission was
%(marginally)
detected only in the
%$\sim$40 ks and $\sim$50 ks
two \cha\ observations carried out in 2000 and 2001
\citep{kulkarni03}.
%These detections were based on rather low significance
%\textbf{[la nostra detection non \`e molto meglio\ldots]} peaks in the
%$Z_2^2$ periodograms, restricted to a range of periods
The periods measured with \cha\ are compatible with the 8 s period
detected in the pulsating tail of the 1979 giant flare and
% The comparison of the measured periods
give a spin-down rate of $\dot{P}=(6.5\pm0.5)\times10^{-11}$ s
s$^{-1}$, a value in the same range as that of the other magnetar
candidates.

In order to measure again the pulsation period and search for
long-term flux and spectral changes, we obtained a $\sim$70 ks long
observation of \src\ with \xmm,  6 years after the latest X-ray
observation. In the following we report on the results of this
recent observation, together with the analysis of two short archival
\xmm\ observations performed in 2000 and 2001.

\section{Observations and data analysis}

\src\ was observed by \xmm\ on 2007 November 10 for about 70 ks. The
field containing \src\ had already been observed by \xmm, with
shorter exposure times,
%in two occasions,
on 2000 July 8 and  on 2001 April 8. In the 2000 observation \src\
was $\sim$6$^{\prime}$ off-axis, while in the other observations it
was on-axis. We concentrate here on the analysis of the data
collected with the EPIC instrument, which is composed by a PN
\citep{struder01short} and two MOS X-ray cameras
\citep{turner01short}, sensitive in the 0.2--15 keV energy range.
Details on the instrument settings (optical blocking filter and
operating mode) for each observation are listed in Table \ref{log}.
For the longest observation we used also the data collected by the
Reflection Grating Spectrometer (RGS, \citealt{rgs01short}), which
worked in parallel to the EPIC instrument and had a net exposure
time of 71 ks for each of its two units (RGS1 and RGS2). This high
resolution spectrometer is sensitive in the 0.35--2.5 keV energy
range.

All the data were processed using the \xmm\ Science Analysis
Software (\textsc{sas} version 8.0.0) and the calibration files
released in August 2007. The standard pattern selection criteria for
the EPIC X--ray events (patterns 0--4 for PN and 0--12 for MOS) were
adopted. The RGS analysis followed the standard selection criteria
as well.\footnote{See\\
http://xmm.esac.esa.int/external/xmm\_user\_support/documentation/sas\_usg/USG/}
Response matrices and ancillary files for each spectrum were
produced using the \textsc{sas} software package and the spectra
were fitted using \textsc{xspec} version 11.3.1. All errors reported
in the following analysis are at 1 $\sigma$.

\begin{table}
\centering \caption{Log of the \xmm\ observations of \src.}
\label{log}
\begin{tabular}{@{}cccccccc}
\hline
Obs. & Date  & Instrument & Mode$^{\rm a}$ & Filter &  Net exposure  \\
%& MOS1 mode$^{\rm a}$/filter & MOS1 exp. & MOS2 mode$^{\rm a}$/filter & MOS2 exp.\\
% & & & (ks) &  &(ks) & & (ks) \\
\hline
%A$^{\rm b}$ & ? 2000 & FF/medium & ? & FF/medium & ? & FF/medium & ?\\
%B & ? 2001  & FF/medium & ? & FF/medium & ? & FF/medium & ?\\
%C & ? 2001  & SW/medium & ? & FF/thick & ? & FF/medium & ?\\
%D & ? Nov 2007  & LW/thick & ? & SW/thick & ? & SW/thick & ?\\
\phantom{$^{\rm b}$}A$^{\rm b}$ & 2000-07-08 & PN & FF & medium & 4.3 ks\\
& & MOS1 & FF & medium & 8.4 ks\\
& & MOS2 & FF & medium & 8.5 ks\\
\cline{1-6}
B & 2001-04-08 & PN & FF & medium & 6.7 ks\\
& & MOS1 & FF & thick & 4.6 ks\\
& & MOS2 & FF & medium & 12.6 ks\\
\cline{3-6}
& & PN & SW & medium & 5.8 ks\\
& & MOS1 & FF & medium & 12.1 ks\\
\cline{1-6}
C & 2007-11-10 & PN & LW & thick & 60.3 ks\\
& & MOS1 & SW & thick & 69.7 ks\\
& & MOS2 & SW & thick & 69.7 ks\\
\hline
\end{tabular}
\begin{list}{}{}
\item[$^{\rm a}$] The time resolution of the operating modes are: PN Full Frame (FF):
73 ms; PN Large Window (LW): 48 ms; PN Small Window (SW): 6 ms; MOS
Full Frame (FF): 2.6 s; MOS Small Window (SW): 0.3 s.
\item[$^{\rm b}$] 6$^{\prime}$ off-axis.
\end{list}
\end{table}

\subsection{Spectral analysis}

%[No MOS data yet! Errors at 90\% c.l. but 1 sigma in Kulkarni. Flux
%in 0.5--10 keV, as in Kulkarni?]

\begin{figure}
\resizebox{\hsize}{!}{\includegraphics[angle=0]{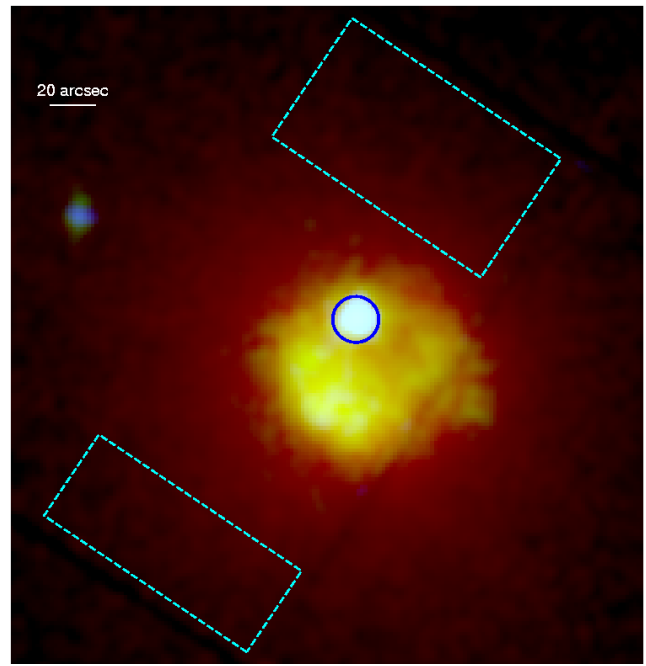}}
\caption{\label{ds9} EPIC PN image of the field of \src\ during the 2007 observation. Photon energy is colour-coded: red corresponds to 0.5--2 keV, green to 2--4 keV, and blue to 4--10 keV. The source and background regions considered for the analysis are overplotted.}
\end{figure}

The spectral analysis of \src\ with \xmm\ is complicated by the
location of this source within the bright SNR N49, whose spatial
extent ($\sim$40$^{\prime\prime}$ radius) is only slightly larger
than the instrumental point-spread function (the 90\% encircled
energy fraction for a point source is $\sim$40$^{\prime\prime}$).
%
%
%
%\src\ is located within the bright SNR N49, whose soft X--ray
%emission heavily contaminates the spectrum of \src. Therefore, t
Rather than attempting to subtract the SNR contribution as a
background component, we included it in the fits with a free
normalization and a fixed spectral shape determined as explained
below. To reduce the contamination from the soft X-ray emission of
N49, we extracted the SGR EPIC spectrum in the 1--10 keV energy
range from a circle of 10$^{\prime\prime}$ radius (this includes
60\% of the point source counts at 5 keV). The background spectrum
was extracted from a region outside the SNR, but in the same CCD as
the SGR (see Figure \ref{ds9}).

%) and modelled the SNR contamination in the spectral fitting of
%the source spectrum.

%\begin{figure}
%\resizebox{\hsize}{!}{\includegraphics[angle=0]{ds9_lin.eps}}
%\caption{\label{ds9} EPIC PN image.}
%\end{figure}

%  and extracted the source spectrum from a small circle
% (10$^{\prime\prime}$ radius) centred at the SGR position (this
% includes 60\% of the point source counts at 5 keV). Despite this,
% a some contamination is still present in the spectral extraction
% region we therefore included the SNR emission in the spectral model.

%spectrum, which is much softer than that of the SGR, is spatially variable.
% Given the relatively small dimensions of the SNR and its spatially
% variable spectrum,
%($\sim$40$^{\prime\prime}$
%radius) in comparison with the instrumental point spread function
%( $\sim$15$^{\prime\prime}$ half energy width) and the spatial
%variability of its spectrum \citep{park03}, the contribution of
%the SNR in the source extraction region cannot be subtracted from
%a SNR region in the vicinity of the point source. We therefore
%extracted the background spectrum from a region outside the SNR
%(but in the same CCD as the SGR) and modelled the SNR
%contamination in the spectral fitting of the source spectrum.

To model the soft and line-rich spectrum of the SNR, we took
advantage of the high-resolution spectra collected by the RGS
instrument during the longest observation. We extracted the first
order spectra from the standard region normally used for point
sources, setting the centre of the SNR as source position
%at the well known SGR position \cite{kulkarni03} and large enough to include ...\% of a point
%source counts.
(such a selection includes most of the photons detected from the
SNR, thanks to its relatively small extension). The RGS spectral
analysis was restricted to the 1--2 keV energy range. To model the
SNR above 2 keV, we extracted a 1--10 keV PN spectrum from a
40$^{\prime\prime}$ circle centred in the middle of the SNR and
fitted it together with the spectra of the two RGS units. Based on
the results of
%high spatial resolution
previous X--ray observations of N49 \citep{park03,bilikova}, we
%started our spectral fitting from
used a model consisting of the sum of two plane-parallel shock
components at different temperatures (\textsc{vpshock} in
\textsc{xspec})
%and a power-law to model the SGR emission,
both corrected for photoelectric absorption (\textsc{phabs} in
\textsc{xspec}). To this we added an absorbed power-law  to account
for the emission from \sgr . Its parameters were fixed at the
best-fit values found with \cha\ : \nh$=$5.6$\times$10$^{21}$
cm$^{-2}$, $\Gamma$=3.06, norm=1.18$\times$10$^{-3}$ s$^{-1}$
cm$^{-2}$ keV$^{-1}$ at 1 keV \citep{kulkarni03}. An overall
normalization factor for each spectrum was also included to account
for the cross-calibration uncertainties between the two RGS and the
EPIC PN.
%Clear narrow emission features in the RGS spectra could not be
%accounted for by the chosen model and therefore required the
%addition of ... gaussian line components, which drastically improved
%the fit (). Since our fit is only phenomenological...
A satisfactory fit ($\chi^2$=2170/935 degrees of freedom,
d.o.f.)\footnote{Although this fit is not statistically acceptable,
a detailed modelling of the SNR emission is beyond the scope of this
paper and so we did not adopt more complex spectral models. } was
obtained with the following parameters:
\nh=(1.3$\pm$0.3)$\times$10$^{21}$ cm$^{-2}$,
%$\Gamma$=7.8,
$kT_1$=0.577$^{+0.002}_{-0.005}$ keV,
$\tau_1$=5.4$^{+1.1}_{-0.8}$$\times$10$^{12}$ s cm$^{-3}$,
$kT_2$=1.10$\pm$0.01 keV, $\tau_2$$>$3$\times$10$^{13}$ s cm$^{-3}$,
Ne/Ne$_{\odot}$=0.66$\pm$0.02, Mg/Mg$_{\odot}$=0.59$\pm$0.01,
Si/Si$_{\odot}$=0.79$\pm$0.01, S/S$_{\odot}$=1.00$\pm$0.04,
Ca/Ca$_{\odot}$=0.6$\pm$0.4, Fe/Fe$_{\odot}$=0.45$\pm$0.02. The
abundances of the other elements are fixed to the Solar values
\citep{anders89}. These parameters are in good agreement with
previous studies of this SNR \citep{park03,bilikova}.
%The flux in the 1--10 keV energy range is 7$\times$10$^{-12}$ \flux; in particular, t
The unabsorbed flux in the 1--10 keV energy is 7.1$\times$10$^{-12}$
\flux for the SNR and 1.6$\times$10$^{-12}$ \flux\ for the SGR.

% contribution to the total flux.

%[better fit with addition of 3 gaussian lines and different
%abundances for 2 pshock components ($\chi^2=1486/922$ d.o.f.).
%Fixing SGR spectrum to Kulkarni par., SNR parameters more similar to
%Park...].

\begin{figure}
\resizebox{\hsize}{!}{\includegraphics[angle=-90]{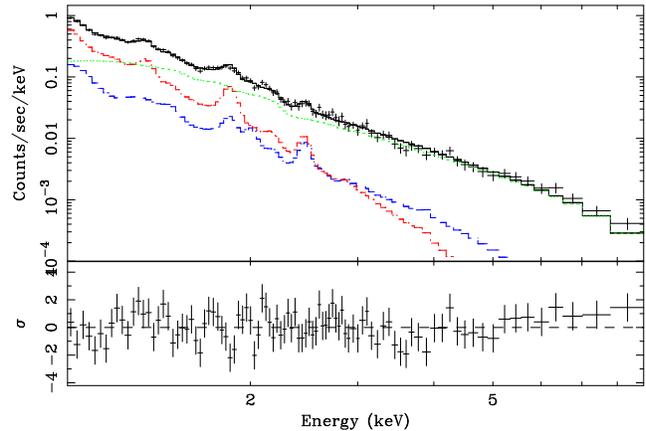}}
\caption{\label{spectrum} EPIC PN spectrum of \src\ collected during
the longest \xmm\ observation in 2007. The model, obtained by a
simultaneous fit of the PN and MOS spectra (see the corresponding
parameters in Table \ref{fits}), is an absorbed power-law (green)
and the sum of two plane-parallel shock components (in blue the
warmer and in red the cooler one) to model the contamination from
the SNR.}
\end{figure}

We then fitted the EPIC spectra extracted from the small region
around \sgr\  with an absorbed power-law\footnote{More complex spectral models, which are usually used to fit magnetar spectra, were not adopted in this case due to the uncertainty of the background subtraction.} plus the model of the SNR
described above. All the SNR model parameters were fixed at their
best-fit values, except for the normalization, in order to properly
account for the unknown intensity of the SNR emission in the source
extraction region.\footnote{We also applied a second normalization
factor to both the SNR and the SGR  models to account for the
cross-calibration uncertainties between the EPIC cameras; the
maximum flux discrepancy we find between the EPIC cameras is lower
than 15\%.} A good fit ($\chi^2=274.1/233$ d.o.f., see Figure
\ref{spectrum}) is obtained with a hydrogen column density \nh\ =
(4.6$^{+0.7}_{-0.5}$) $\times 10^{21}$ cm$^{-2}$ and a photon index
$\Gamma$=3.27$^{+0.07}_{-0.04}$. The lack of systematic residuals in
correspondence with the SNR brightest spectral lines (see Figure
\ref{spectrum}) indicates that the SNR contamination is sufficiently
well modelled. The 1--10 keV (unabsorbed) flux of the power-law
component is (1.25$^{+0.05}_{-0.02})\times$10$^{-12}$ \flux . This
corresponds to a luminosity of 4.3$\times$10$^{35}$ \lum\ for a
distance  of 55 kpc.

%A slightly better fit ($\chi^2=91.0/90$
%d.o.f.) is obtained adding a blackbody component to the SGR spectral
%model; the corresponding best-fit parameters are: \nh\ $<$3$\times
%10^{21}$ cm$^{-2}$, $\Gamma$=2.4$^{+0.5}_{-0.2}$,
%kT=0.37$^{+0.03}_{-0.07}$ keV. Assuming a source distance of 50 kpc,
%the 1--10 keV luminosity of the power-law component is
%1.6$^{+0.7}_{-0.3}\times 10^{35}$ \lum\ and the blackbody emitting
%region has a radius of 7.6$^{+4.4}_{-0.8}$ km.

Since no time variability is expected in the  SNR contribution
%amination is expected to be constant with time,
%and the source extraction region is the same,
we can fit the SGR spectra of the older EPIC observations with the
model described above, keeping the SNR model normalization fixed at
the value obtained in the longest observation. The best-fit
parameters for an absorbed power-law model are reported in Table
\ref{fits}.
No
%significant
spectral variability is detected and
%we can exclude
significant ($>$3$\sigma$) flux variations larger than $\sim$50\%
among the different \xmm\ observations can be excluded.

\begin{table*}
\begin{minipage}{12cm}
\centering \caption{Best-fit spectral parameters for the three EPIC
observations (see Table \ref{log}) of \src\ in the 1--10 keV energy
range. The spectral model consists of a fixed component modelling
the SNR contamination (see text for details) and an absorbed
power-law model for the SGR emission.} \label{fits}
\begin{tabular}{@{}cccccc}
\hline
Observation & SNR norm.$^{\mathrm{a}}$ & \nh\ & $\Gamma$ & Flux$^{\mathrm{b}}$ & $\chi^{2}_{r}$ (d.o.f.)\\
& & (10$^{21}$ cm$^{-2}$) & & (10$^{-12}$ \flux) & \\
\hline
A & 0.117 (fixed) & 3.8$^{+0.2}_{-1.6}$ & 3.1$^{+0.4}_{-0.1}$ & 1.3$^{+0.2}_{-0.1}$ & 1.04 (65) \\
B & 0.117 (fixed) & 5.3$^{+0.6}_{-1.1}$ & 3.3$^{+0.1}_{-0.2}$ &
1.3$\pm$0.1 & 0.88 (166) \\
C & 0.117$^{+0.002}_{-0.005}$ & 4.6$^{+0.7}_{-0.5}$ & 3.27$^{+0.07}_{-0.04}$ & 1.25$^{+0.05}_{-0.02}$ & 1.18 (233) \\
\hline
\end{tabular}

 \medskip
\begin{list}{}{}

%\item[$^{\mathrm{a}}$] The blackbody radii and temperatures are at infinity and assuming a distance of \ldots\ kpc.
\item[$^{\mathrm{a}}$] Normalization factor applied to the best-fit model of the RGS and PN spectrum of the whole SNR.
\item[$^{\mathrm{b}}$] Unabsorbed flux in the 1--10 keV range.
\end{list}
\end{minipage}
\end{table*}

\subsection{Timing analysis}

\begin{figure}
\resizebox{\hsize}{!}{\includegraphics[angle=0]{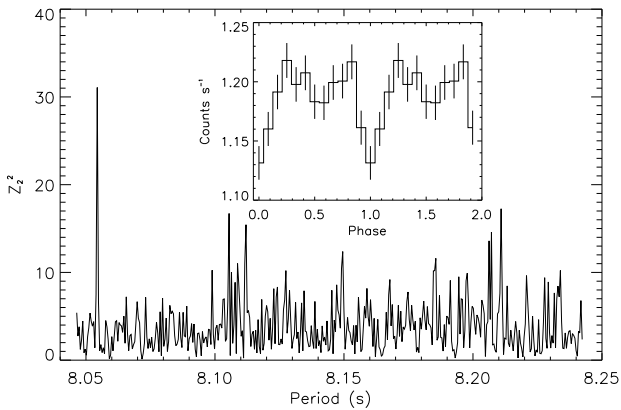}}
\caption{\label{timing}  Z$^2_2$ diagram of the long \xmm\
observation (PN and MOS data in the 0.65--12 keV energy range) of
\src\ in the range used for the period search (see Section 2.2).
%as in \citet{kulkarni03}.
The peak at 8.0544 s is significant at
3.2$\sigma$. \emph{Inset:} The corresponding EPIC pulse profile
(0.65--12 keV, not background subtracted).}
\end{figure}

For the timing analysis the photon arrival times were converted to
the solar system barycenter with the \textsc{barycen} \textsc{sas}
task.
% and the source photons were accumulated in the
We used the same extraction region adopted for the spectra, but we
performed the analysis in the 0.65--12 keV energy band to
%further minimize contamination by the supernova remnant.\\
optimize the signal to noise ratio.
%\indent
The pulse periods measured with \cha\ using the $Z_2^2$ test were
%The pulse period of \src\ was measured with \cha\ by at
8.0436(2) s on 2000 January 4 and 8.0470(2) s on 2001 August 31
\citep{kulkarni03}.
%Based on these  results, w
We searched for pulsations in the period range 8.0464--8.2423 s,
selected by considering the 3$\sigma$ lower limit on the most recent
spin period value reported by \citet{kulkarni03} and extrapolating
to the epoch of the \xmm\ observation under the assumption of a
(conservative) period derivative of $0\leq\dot{P}\leq10^{-9}$ s
s$^{-1}$. In Figure \ref{timing} we show the $Z_2^2$ periodogram
obtained by using the combined events from the PN and MOS cameras.
Taking into account the number of searched periods (419), the peak
value of 31.08 (chance probability of $\sim$$3\times10^{-6}$) at
$\sim$8.0544 s corresponds to a significance of $\sim$99.88\% (that
is a 3.2$\sigma$ detection).\\
\indent To better estimate the period, we used an epoch folding
technique and fitted the peak in the $\chi^2$ versus trial period
distribution as described in \citet{leahy87}. This led to
$P=8.0544\pm0.0002$ s. The resulting EPIC folded lightcurve is shown
in Figure \ref{timing}. The pulsed fraction, defined as
$(C_{\rm{max}}-C_{\rm{min}})/(C_{\rm{max}}+C_{\rm{min}})$, where
$C_{\rm{max}}$ and $C_{\rm{min}}$ are the background-subtracted
count rates at the peak and at the minimum, is $(13.6\pm0.9)$\% in
the 0.65--12 keV energy range.\footnote{ The SNR contamination,
evaluated from the model described in Section 2.1, is also included
in the background. The error in the pulsed fraction does not include
the systematic uncertainty due to the SNR contamination.} We
repeated a similar analysis on the data-sets from the two short
\xmm\ observations carried out in 2000 and 2001 (see Table
\ref{log}), but, due to the paucity of photons,
we did not detect any significant periodicity.\\
\indent  Considering also the two periods measured by \cha\
\citep{kulkarni03}, a linear fit to the period evolution of \src\ is
unacceptable ($\chi^2_{r}$ of 15.6 for 1 degree of freedom).
%\textbf{[probability = $\mathbf{7.832\times10^{-5}}$]};
%$\dot{P}\simeq4.2\times10^{-11}$ s s$^{-1}$).
On the other hand, the period derivative inferred from our period
measurement and the most recent \cha\ detection is
$(3.8\pm0.1)\times10^{-11}$ s s$^{-1}$. Comparison with the value
derived from the two \cha\ measurements
($\dot{P}=(6.5\pm0.5)\times10^{-11}$ s s$^{-1}$) indicates a
significant decrease of the spin-down rate.

% We note that the most recent period measurements are all above the
% average trend, suggesting an increase of the spin-down rate.
% However, the period derivative obtained fitting only the 7 periods
% measured after 2004 ($\dot{P}=(3.0\pm1.4)\times10^{-11}$ s
% s$^{-1}$) does not indicate a statistically significant increase.

%\begin{figure}
%\resizebox{\hsize}{!}{\includegraphics[angle=0]{z2.eps}}
%\caption{\label{z2} Z$^2_2$ diagram of the long \xmm\ observation
%(PN and MOS data in the 0.65--12 keV energy range) of \src\ in the
%same period range as in \citet{kulkarni03}. The peak at 8.0543848 s
%is significant at ...$\sigma$.}
%\end{figure}

%\begin{figure}
%\resizebox{\hsize}{!}{\includegraphics[angle=0]{fig1.eps}}
%\caption{\label{history} Long term light-curve of \src\ based on data from different satellites (updated from \citealt{mereghetti06}). The vertical dashed lines indicate the onset of the two burst-active periods of the source. The inset shows in detail the light-curve around the 2008 reactivation, using also \cxo\ data from \citet{tiengo08atel1559} and \citet{woods08atel1564}. The down-arrows indicate upper limits at 3\,$\sigma$ confidence level.}
%\end{figure}

%\begin{table}
%\centering
%\caption{Journal of the 2008 \swift/XRT observations. The observation sequence number is composed of 00312579 followed by the three digit segment number given here (e.g. 00312579001).}
%\label{log}
%\begin{tabular}{@{}cccccc}
%\hline
%Obs. & Date & \multicolumn{2}{c}{Start/End time (UT)} & Exposure$^{\rm a}$ & Count rate\\
% & mm-dd & \multicolumn{2}{c}{hh:mm:ss} & (ks) & (counts s$^{-1}$)\\
%\hline
%001 & 05-28 & 12:58:14 & 13:31:27 & 2.0 & $0.067 \pm 0.003$\\
%002 & 05-29 & 14:47:17 & 16:30:39 & 2.0 & $0.015 \pm 0.003$\\
%\hline
%\end{tabular}
%\begin{list}{}{}
%\item[$^{\rm a}$] The exposure time is usually spread over several snapshots (single continuous pointings at the target) during each observation.
%\item[$^{\rm b}$] Upper limit at 3\,$\sigma$ confidence level \citep[following][]{gehrels86}.
%\end{list}
%\end{table}

\section{Discussion and conclusions}

Due to its location in the LMC, \src\ is one of the magnetars less
frequently observed. In addition to its larger distance with respect
to Galactic magnetars, the analysis of its persistent X--ray
emission is complicated by the bright soft X--ray emission of the
N49 SNR. With a deep \xmm\ observation performed in 2007, we could
measure the pulsation period of \src\, which was previously detected
only in the pulsating tail following the 1979 March 5 giant flare
\citep{mazets79} and in two \cha\ observations taken in 2000 and
2001 \citep{kulkarni03}. The pulsation profile (shown in Figure
\ref{timing}) is double-peaked and the pulsed fraction is
(13.6$\pm$0.9)\%. Although, to our knowledge, the pulse profile of
\src\ was never published before, these might be permanent
properties of this source since a non-sinusoidal modulation and a
pulsed fraction around 10\% were also reported for the \cha\ data
\citep{kulkarni03}.

Although the period measurements of \src\ are very sparse, the value
we derived shows, for the first time, a significant decrease in the
spin-down rate of this source. In the magnetar model, a reduction of
the spin-down rate can be interpreted as an indication of a more
relaxed state of the twisted magnetosphere and should be associated
to a low rate of bursting activity, a spectral softening and a
decrease of the persistent X--ray luminosity \citep{tlk02}. This
behavior is sometimes observed in magnetar candidates (see, e.g.,
\citealt{mte05}), but some exceptions have been found (see, e.g.,
\citealt{gavriil04}). In the case of \src, the bursting activity is
indeed very low, since no bursts have been detected since 1983.
% is not compatible with the spin-down trend obtained with the previous observations.
%This luminosity is comparable to that of the most active SGRs
%(SGR\,1806--20 and SGR\,1900+14), even if the last burst from \src\
%was detected more than 20 years ago.
However, we note that some bursts might have been missed due to its
large distance and its location in a sky region not frequently
monitored by $\gamma$-ray instruments.

The X--ray luminosity measured in 2007 is $\sim30$\% lower (and the
spectrum slightly softer) than that reported from the analysis of
\cha\ data taken in 2000 and 2001 \citep{kulkarni03}. However, due
to the different characteristics of the instruments and additional
uncertainties due to the presence of contamination from the SNR
diffuse emission, we consider these changes well within the
systematic uncertainties. Using the two shorter \xmm\ observations,
taken almost simultaneously with the \cha\ ones, we can instead
extract the spectrum and model the SNR emission in the same way as
we did for the 2007 observation. In this case we are dominated by
statistical errors and no significant changes in the spectral shape
and source flux are detected.

The luminosity of \src, which can be well determined thanks to its
accurately known distance, is higher than that of most magnetar
candidates (see e.g. \citealt{durant06}). Since this high luminosity
has not substantially varied for at least several years,
%showing that probably this bright state
it is probably a long-lasting property of this magnetar rather than
a transient bright state related to its past bursting activity,
culminating with the giant flare of 1979 March 5.
%\sgr\ should therefore be considered a persistent magnetar rather than a
%transient one;
This behavior is radically different from the one displayed, as an
example, by SGR~1627--41 that, after two distinct periods of
bursting activity, rapidly decreased its persistent X--ray
luminosity down to $\sim$10$^{33}$ \lum\ \citep{eiz08short}. While
it is unclear whether this behavior is related to intrinsic
differences between the sources (e.g. the magnetic field) or
different evolutionary stages, this seems to support the emerging
trend of separating the magnetars into transient and persistent
objects, rather than in AXPs and SGRs.
%The large amount of data now available on magnetars have shown that
%the main distinction among these objects, possibly due to intrinsic
%or evolutionary differences, might be their transient or persistent
%behavior, rather than the traditional distinction between AXPs and
%SGRs. In fact, many AXPs have been observed to emit SGR-like bursts
%and the defining characteristics of the AXPs are also fulfilled by
%all the persistent counterparts of SGRs. On the other hand, there is
%a clear dichotomy between persistent sources, whose luminosity can
%vary of only a factor few, and the transient ones, which, during
%their outbursts, can increase their luminosity by two orders of
%magnitude.
%%. which are observed to be bright (L$_{\rm X}$$\sim$10$^{35}$ \lum)
%%X--ray sources for a limited fraction of their lifetime and the
%%persistent ones,
%%, determined by the eventual presence of bursting activity.
In this framework, \src\ should therefore be considered a member of
the persistent magnetars group.

\section*{acknowledgements}
We acknowledge the partial support from ASI (ASI/INAF contracts
I/088/06/0 and AAE~TH-058). PE thanks the Osio Sotto city council
for support with a G.~Petrocchi fellowship. SZ acknowledges support
from STFC. NR is supported by a NWO Veni Fellowship. DG acknowledges
the CNES for financial support.

\bibliographystyle{mn2e}
\bibliography{biblio}
\bsp

\label{lastpage}

\end{document}